# Phase-locked beam splitting with continuously tunable intensity ratio using stochastic phase masks

Melvin Redon[1,*], Titouan Gadeyne[2], Gaspard Beaufort[1], Saga Westerberg[1], Vartika Vishnoi[2], Peter Smorenburg[1,3], Anne-Lise Viotti[1], Thierry Ruchon[2] and Cord L. Arnold[1]

[1]Department of Physics, Lund University, P.O. Box 118, SE-22100 Lund, Sweden
[2]Université Paris-Saclay, CEA, LIDYL, 91191 Gif-sur-Yvette, France
[3]ASML Research, ASML Netherlands B.V., 5504 DR Veldhoven, The Netherlands
[*]Author to whom any correspondence should be addressed.
E-mail: melvin.redon@fysik.lu.se

**Abstract.** We introduce a common-path beam splitter based on a randomly partitioned phase mask displayed on a spatial light modulator (SLM). The SLM aperture is divided into square patches, each randomly assigned to one of three phase ramps that steer light to three independent foci: two phase-locked beams and a dump that takes up the surplus power. We observe the beams in the focal plane of a lens. The fractions of patches assigned to each ramp set the focal intensities, with the focus positions controlled by the ramp slopes. We first demonstrate experimentally that the intensity ratio between the two foci can be tuned from 0 to 1 without distorting either of the spatial modes, and with their relative phase constant to within a few tens of mrad. We then derive a closed-form expression for the ensemble-averaged focal intensity, which decomposes into coherent foci of the desired pattern and a speckle background whose shape and amplitude are set by the patch size. The model is validated against experiment and against numerical Fraunhofer propagation, with quantitative agreement over about three orders of magnitude in intensity. We also identify the patch size as the central design parameter and characterize its experimental bounds.

## 1 Introduction

Spatial interferometry provides a direct route to measuring how a phase evolves with a control parameter: two spatially separated, phase-locked sources are generated, one is perturbed, and the resulting fringe shift in the interference pattern gives access to the induced phase difference. However, any relative path-length drift between the two sources shifts the fringes independently of the effect under study, which makes two-arm geometries sensitive to mechanical vibrations and thermal drift. Producing both sources from a single beam suppresses this problem naturally: residual fluctuations are shared by both arms and cancel in the fringe pattern. Common-path interferometers have found use across a broad range of applications, from optical coherence tomography [1] to two-dimensional electronic spectroscopy [2] and the characterization of nonlinear optical processes including high-order harmonic generation (HHG) [3,4]. For the latter, it has been shown that two harmonic sources driven by the same laser are mutually coherent [5], and pairs of phase-locked harmonic sources have since served for interferometric measurements of the atomic dipole phase [6], extreme-ultraviolet (XUV) Fourier-transform spectroscopy [7,8], and attosecond interferometry of atoms, molecules and solids [9–12]. In several of these applications it is desirable not only to phase-lock the two sources but also to control their relative intensity, so that, for example, an intense pump and a weak probe can be generated from the same beam.

A natural way to realize such a common-path geometry is to split a single collimated beam into two spatially separated foci at the focal plane of a lens. Several optical elements have been used to this end. For instance, a $\pi$-step phase plate imprints opposite phases on the two halves of the beam, which interfere constructively at two symmetric off-axis positions [13]. A binary $0-\pi$ grating achieves a similar result through diffraction [14]. These elements belong to the wider family of multi-beam phase gratings, from binary Dammann gratings [15] to continuous profiles of optimal efficiency [16–18], all of which fix the intensity ratios in the grating profile. A phase-only spatial light modulator (SLM) can be a more flexible alternative. It is a pixelated device that imparts a programmable, spatially varying phase pattern, or mask, onto a reflected beam without affecting its amplitude [19, 20]. Tross *et al.* showed that a binary checkerboard mask, assigning alternating pixels to one of two prism (linear-ramp) phases, steers light into two symmetric foci, each fed by one interleaved subset of pixels [4]. All of these approaches share a fundamental limitation: since both foci are created simultaneously at the focal plane by a single incoming beam, there is no independent optical path upstream of the focus through which an attenuator could be inserted to control the intensity of each source separately. The intensity ratio is therefore fixed by the mask geometry, at 50/50 in all cases above. Only recently was independent intensity control achieved, while preserving the spatial modes, using a pair of birefringent wedges (TWINS) [2, 3], although at the cost of a spatially fixed splitting geometry.

In the present work, we lift the central limitation of the checkerboard SLM scheme, namely its fixed 50/50 intensity ratio, by replacing the periodic mask with a randomly partitioned phase mask. The relative intensities of the beams then become continuously and independently tunable through the patch fractions alone (the share of the SLM's patches assigned to each channel), while the common-path geometry and the spatial modes of the foci are preserved. The concept is sketched in Fig. 1: the SLM aperture is tiled into patches, and the fraction of patches assigned to each channel sets the corresponding focal intensity. Random spatial multiplexing of separate functions on a single aperture was first demonstrated by Caulfield using randomly interleaved holographic masks [21]. Davis and Cottrell later applied this idea to phase-only SLMs, showing that multiple functions can be multiplexed with independently controllable relative densities [22, 23]. The principle was subsequently refined for holographic optical trapping [24–27] and for complex amplitude encoding [28–31]. We adapt this principle to the design of a three-channel common-path beam splitter and provide a closed-form description of the resulting focal-plane intensity, identifying the patch size as the central design parameter and quantifying its experimental bounds. The method offers a non-iterative alternative to Gerchberg-Saxton holography [32–36], with a closed-form mask formula and a predictable speckle structure.

The paper is organized as follows. Section 2 introduces the experimental setup, the stochastic three-channel mask, and the parameters that set the patch fractions. Section 3 demonstrates the key properties of the splitter experimentally: continuous, mode-preserving tuning of the intensity ratio at constant relative phase. Section 4 develops the closed-form theory of the focal-plane intensity. Section 5 validates it against numerical propagation and experiment. Section 6 treats the patch size as the quantity the user has to choose, and establishes its bounds.

# 2 Stochastic three-channel mask

## 2.1 Setup and notation

The setup is sketched in Fig. 1. The experiments use a reflective phase-only liquid-crystal-on-silicon (LCoS) SLM (Santec SLM-300, 1920×1200 pixels, pixel pitch $p = 8$ µm, phase modulation depth $2\pi$) at the working wavelength $\lambda = 1030$ nm. The SLM is illuminated at near-normal incidence by a collimated Gaussian beam of $1/e^2$ intensity radius $w_{\mathrm{SLM}} = 3$ mm, well within the active aperture, with its polarization aligned to the modulation axis of the liquid crystal. The field after passing the SLM is $E_{\mathrm{SLM}}(X, Y) = A_{\mathrm{SLM}}(X, Y)\, e^{i\phi(X,Y)}$, with Gaussian envelope $A_{\mathrm{SLM}}(X, Y) = A_0 \exp[-(X^2 + Y^2)/w^2_{\mathrm{SLM}}]$ and $\phi(X, Y)$ the phase applied by the SLM. The real beam departs slightly from this ideal envelope (Sec. 5), but since the same envelope multiplies every channel, such aberrations reshape all foci identically and leave their relative intensities unchanged. A lens of focal length $f = 200$ mm placed after the SLM performs the Fraunhofer transform of the reflected field onto the camera, giving the focal field

$$E(x, y) \;\propto\; \frac{1}{i\lambda f} \iint E_{\mathrm{SLM}}(X, Y)\, \exp\!\left[-i\frac{2\pi}{\lambda f}\,(Xx + Yy)\right] dX\, dY \tag{1}$$

with SLM coordinates $(X, Y)$ and focal-plane coordinates $(x, y)$, and fields are normalized so that $|E|^2$ is an intensity, i.e. $\iint |E|^2\, dx\, dy = P_{\mathrm{tot}}$, where $P_{\mathrm{tot}}$ is the total power reaching the focal plane. For an unshaped beam, the diffraction-limited focus has a $1/e^2$ intensity radius $w_0 = \lambda f/(\pi w_{\mathrm{SLM}}) \approx 20$ µm.

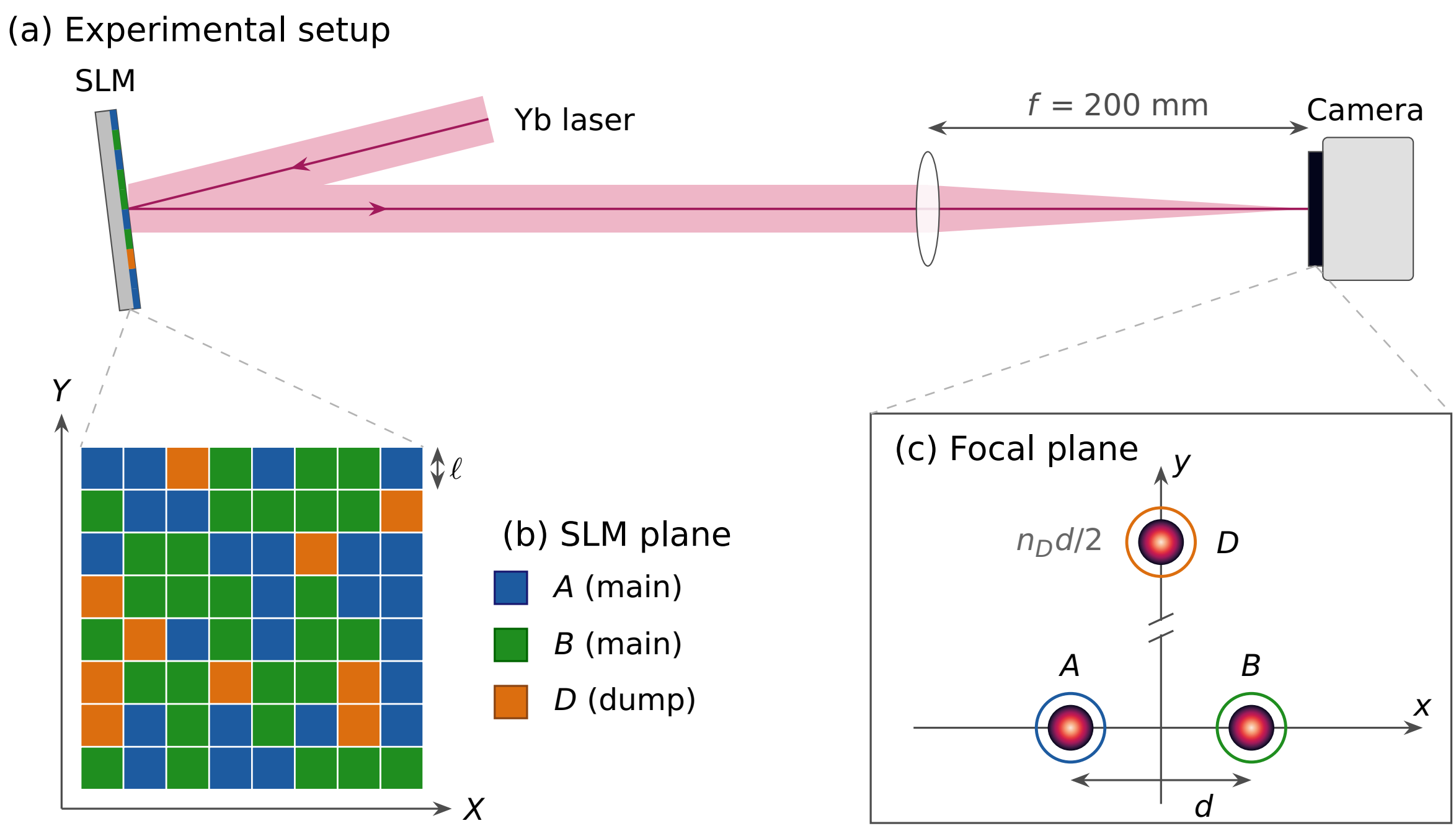

**Figure 1:** Experimental setup and stochastic three-channel mask. (a) A collimated Ytterbium (Yb) laser beam ($\lambda = 1030$ nm) illuminates the phase-only SLM at near-normal incidence, and a lens of focal length $f = 200$ mm performs the Fraunhofer transform onto the camera, placed in its focal plane. (b) Example of a mask realization on the SLM: each patch of side length $\ell$ is randomly assigned to one of the three channels $A$, $B$, $D$, with relative fractions $\xi_A, \xi_B, \xi_D$. (c) Corresponding focal-plane layout on the camera, each focus circled in the color of its channel. The main foci $A$ and $B$ lie on the $x$ axis, separated by $d$, and the dump focus $D$ is deflected along $y$ to $n_D\, d/2$, outside the region of interest (the $y$ axis is not to scale).

A global carrier phase ramp is added to every mask displayed in this work, so that all shaped orders are deflected away from the light that leaves the SLM unshaped. The residual light acquires no SLM phase, including no ramp, and therefore stays in the specular direction, apart from the diffraction orders of the pixel grid, which land $\lambda f/p \approx 26$ mm off axis, far outside the field of view. Focal-plane coordinates below are referred to the deflected frame, and the residual specular spot is the zeroth order quantified in Sec. 6.2.

## 2.2 Three-channel partition

The SLM aperture is divided into a regular grid of square patches of side length $\ell = s\, p$ [Fig. 1(b)], where $s$ is an integer setting the patch size in pixels. Each patch is independently and randomly assigned to one of three *channels*: two main channels $A$ and $B$, directing light into two different focal spots, and a dump channel $D$, which takes up the surplus power. The assignment probabilities, equal to the mean patch fractions, are denoted $\xi_A$, $\xi_B$, $\xi_D$, and their sum is equal to one [Fig. 1(b)]. Three channels are the minimum needed to set the intensities of the two beams independently. With two channels the fractions are complementary, $\xi_A + \xi_B = 1$, so their ratio can still be tuned, but attenuating one focus necessarily brightens the other. The dump breaks that coupling. Light is diffracted at the phase discontinuities between neighboring patches assigned to different channels. Since the assignment is random, the diffracted light does not add up in any focus but forms a speckle halo. The focal-plane intensity $I_{\text{tot}}(x, y)$ therefore splits into a coherent part $I_{\text{coh}}(x, y)$, which forms the three foci, and a halo part $I_{\text{halo}}(x, y)$, i.e. a diffuse speckle background surrounding each focus whose shape is set by the patch geometry, a well-known property of random diffusers [37, 38]. Section 4 derives $I_{\text{coh}}(x, y)$ and $I_{\text{halo}}(x, y)$ explicitly.

Each channel $c$ carries a linear phase ramp. The two main channels are tilted in opposite directions along $X$, and the dump along $Y$ with a slope steeper by a factor $n_D > 1$ that displaces it well outside the region of interest, with phases

$$\phi_A = +k_t X, \quad \phi_B = -k_t X, \quad \phi_D = n_D\, k_t\, Y. \tag{2}$$

The phase displayed on the SLM assigns one of these three ramps to each patch, $\phi(X, Y) = \sum_c \mathcal{I}_c(X, Y)\, \phi_c(X, Y)$. Here $\mathcal{I}_c(X, Y)$ is the indicator function of channel $c \in \{A, B, D\}$: it equals 1 on patches assigned to $c$ and 0 elsewhere. The indicators are non-overlapping and sum to one at every point. The complex mask $M$ therefore splits into a sum over the three ramps,

$$M(X, Y) \equiv e^{i\phi(X,Y)} = \sum_c \mathcal{I}_c(X, Y)\, e^{i\phi_c(X,Y)}, \tag{3}$$

so that $E_{\text{SLM}} = A_{\text{SLM}}\, M$, according to Eq. (3).

Channels $A$ and $B$ are the two *main* beams, separated along $x$ by the distance $d$, set by the transverse wavevector $k_t = \pi d/(\lambda f)$. The three foci therefore sit at positions $(x_c, y_c)$, namely $(\pm d/2, 0)$ for the main channels and $(0, n_D\, d/2)$ for the dump [Fig. 1(c)]. We use $d = 100$ µm and $n_D = 10$ throughout this paper, which places the dump ten times farther off axis than the main foci.

### 2.3 Setting the patch fractions

The three fractions $\xi_A$, $\xi_B$ and $\xi_D = 1 - \xi_A - \xi_B$ must be chosen to achieve the desired intensities in the two main foci and the dump.

**No dump.** We begin by splitting the SLM patches into the two main channels $A$ and $B$ only (no dump). First, we recall that some of the total power $P_{\text{tot}}$ always contributes to the halo described above, diffracted at the discontinuities between $A$ and $B$ patches. We call $P_{\text{coh}}$ the remaining coherent power and introduce a first parameter $\alpha \in [0, 1]$ which fixes how this power is split between the two main channels (for $\alpha = 0$ all the power is sent to $A$, for $\alpha = 1/2$ the two are balanced),

$$P_A = P_{\text{coh}}\, (1 - \alpha), \qquad P_B = P_{\text{coh}}\, \alpha. \tag{4}$$

As will be derived in Sec. 4, the power in a coherent focus $c$ is $P_c = \xi_c^2\, P_{\text{tot}}$. Equating $P_A = P_{\text{coh}}(1 - \alpha) = \xi_A^2 P_{\text{tot}}$ and $P_B = P_{\text{coh}}\, \alpha = \xi_B^2 P_{\text{tot}}$, with $\xi_A + \xi_B = 1$, gives

$$\xi_A = \frac{\sqrt{1-\alpha}}{\sqrt{1-\alpha} + \sqrt{\alpha}}, \qquad \xi_B = \frac{\sqrt{\alpha}}{\sqrt{1-\alpha} + \sqrt{\alpha}}, \tag{5}$$

and $P_{\text{coh}} = P_{\text{tot}}/(1 + 2\sqrt{\alpha(1-\alpha)})$, the remainder $P_{\text{tot}} - P_{\text{coh}} = 2\xi_A \xi_B\, P_{\text{tot}}$ going into the halo. The coherent power therefore depends on how evenly the patches are split between the two channels, and is lowest for $\alpha = 1/2$.

**With dump.** Then, two other parameters $\beta_A$ (resp. $\beta_B$) are introduced to control the attenuation of focus $A$ (resp. $B$): $\beta_A = 0$ leaves focus $A$ untouched and $\beta_A = 1$ extinguishes it. This attenuation proceeds by two mechanisms: (i) part of the $A$ patches are reassigned to the $D$ channel, but also (ii) the new discontinuities between $A$ and $D$ patches diffract part of the light into the halo. Note that the coherent power in the $B$ channel should not be affected by $\beta_A$, and vice versa. Ultimately, the powers in the two main foci should obey

$$P_A = P_{\text{coh}}\,(1-\alpha)(1-\beta_A), \qquad P_B = P_{\text{coh}}\,\alpha\,(1-\beta_B), \tag{6}$$

where $P_{\text{coh}}$ keeps its no-dump value, fixed by $\alpha$ alone, and serves as the reference for the attenuation. With $P_c = \xi_c^2\,P_{\text{tot}}$, the fractions become

$$\xi_A = \frac{\sqrt{(1-\alpha)(1-\beta_A)}}{\sqrt{1-\alpha}+\sqrt{\alpha}}, \qquad \xi_B = \frac{\sqrt{\alpha(1-\beta_B)}}{\sqrt{1-\alpha}+\sqrt{\alpha}}, \qquad \xi_D = 1-\xi_A-\xi_B. \tag{7}$$

In practice, each patch is first assigned to the $A$ or $B$ side in the proportion $\sqrt{1-\alpha} : \sqrt{\alpha}$, set by $\alpha$ alone, and a patch on the $A$ side is then sent to the dump with probability $1-\sqrt{1-\beta_A}$, likewise for $B$. The dump then receives $P_D = \xi_D^2\,P_{\text{tot}}$. With these fractions set, the power (or peak-intensity) ratio between the two main foci is

$$\frac{P_A}{P_B} = \frac{\xi_A^2}{\xi_B^2} = \frac{1-\alpha}{\alpha}\cdot\frac{1-\beta_A}{1-\beta_B}. \tag{8}$$

## 3 Experimental demonstration

Before developing the theory, we demonstrate the two properties that motivate the method: (i) the intensity ratio between the two main foci can be tuned continuously while their spatial modes are preserved, and (ii) the relative phase of the pair does not move while it is tuned.

To verify this experimentally, we set $\alpha = 1/2$ and $\beta_B = 0$, and scan $\beta_A$ from 0 to 1, for which Eq. (8) predicts $I_A/I_B = 1-\beta_A$. For each setting we record the focal-plane image and integrate the power in the two main foci and in the dump [Fig. 2(a)]. Focus $A$ decreases linearly with $\beta_A$, as predicted, while focus $B$ stays constant: the patches taken from $A$ go to the dump, not to $B$. The dump, however, does not grow linearly but as $(1-\sqrt{1-\beta_A})^2$, because the power in a focus scales as the square of its patch fraction. Part of the power taken from $A$ therefore never reaches the dump and is lost into the halo (Sec. 2.3 and Sec. 1 of the supplementary material). The focal-plane images [Fig. 2(c)–(e)] confirm that focus $A$ fades continuously while focus $B$ keeps both its intensity and its shape.

The relative phase between the two foci is preserved across the scan. Translating the camera away from the focal plane lets the two diverging beams overlap and interfere, and the fringe position tracks the relative phase. We extract it from the Fourier component of the recorded pattern at the fringe spatial frequency. The relative phase stays constant to within 48 mrad rms across the scan [Fig. 2(b)], so attenuating one channel does not shift the phase of the pair. The error bar is set by the signal-to-noise ratio of the fringe peak, and therefore grows as the fringes lose contrast. Beyond $\beta_A = 0.7$ focus $A$ is too weak for the fringes to hold contrast, and those points (open markers) are excluded from the rms deviation quoted above.

## 4 Focal-plane intensity

We now derive $I_{\text{coh}}(x, y)$ and $I_{\text{halo}}(x, y)$, the coherent and halo parts of the focal-plane intensity induced by the mask of Eq. (3). The treatment adapts the speckle analysis originally developed for a random absorbing screen of binary (0/1) amplitude transmittance, later extended to phase screens [37,38], to our multi-channel phase partition.

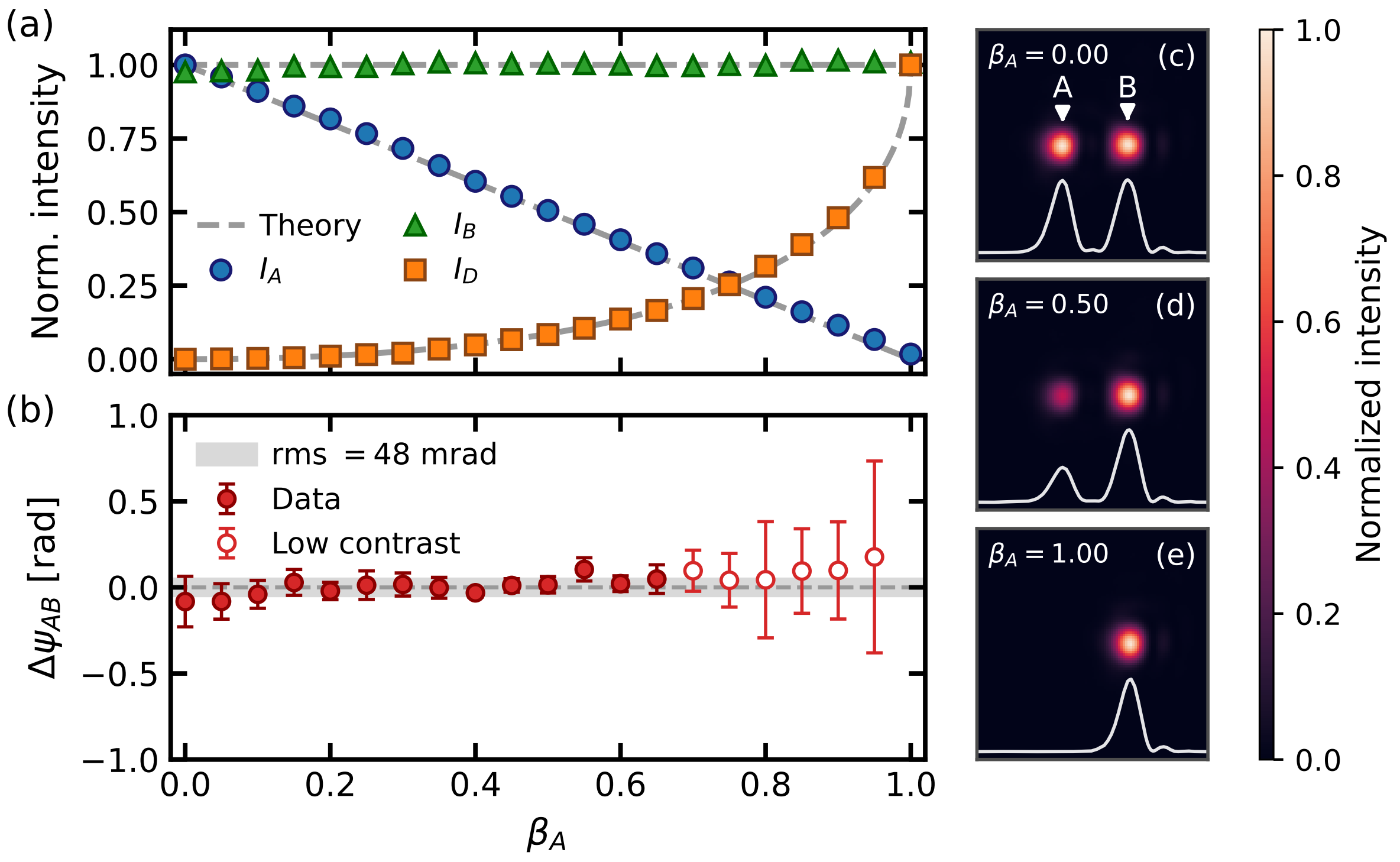

**Figure 2:** Experimental demonstration of independent intensity control. (a) Measured power in focus $A$, focus $B$ and the dump $D$ as a function of the attenuation $\beta_A$, with $\alpha = 1/2$ and $\beta_B = 0$, normalized to $I_A(0)$, the mean of $I_B$ and $I_D(1)$ respectively. Dashed lines: predicted $1 - \beta_A$, 1 and $(1 - \sqrt{1 - \beta_A})^2$. (b) Relative phase $\Delta\psi_{AB}$ between the two foci over the same scan, extracted from the interference fringes recorded away from the focal plane. It remains constant to within $\sim$ 48 mrad rms. The uncertainty grows as $\beta_A \to 1$, where focus $A$ vanishes and the fringe contrast drops. (c)–(e) Focal-plane images of the two main foci at $\beta_A = 0$, 0.5, 1, with foci $A$ and $B$ identified by arrows in (c) and the $y = 0$ lineout traced in white on a linear scale: focus $A$ is continuously suppressed while focus $B$ remains unchanged.

### 4.1 Mean and fluctuation

Each realization of the random partition yields a different mask $M(X,Y)$, and through Eq. (1) a different focal field. Since a single mask is only one random draw, we ask what the focal intensity looks like on average over many such draws. At a given point of the aperture the patch belongs to $A$, $B$ or $D$ with probabilities $\xi_A, \xi_B, \xi_D$, so averaging over realizations replaces the one ramp actually displayed there by the weighted superposition of all three. We write $\langle\cdot\rangle$ for that average and split the mask into its mean and the deviation from it,

$$M = \langle M\rangle + \delta M, \qquad \langle \delta M\rangle = 0, \tag{9}$$

where the second relation holds by construction. The illumination $A_{\mathrm{SLM}}$ is the same for every realization and the Fraunhofer transform of Eq. (1) is linear, so this split carries through to the focal plane: inserting $A_{\mathrm{SLM}}\,\langle M\rangle$ and $A_{\mathrm{SLM}}\,\delta M$ in place of $E_{\mathrm{SLM}}$ in Eq. (1) gives $\langle E\rangle$ and $\delta E$, respectively, with

$$E = \langle E\rangle + \delta E, \qquad \langle \delta E\rangle = 0, \tag{10}$$

so that

$$|E|^2 = |\,\langle E\rangle\,|^2 + 2\,\mathrm{Re}\big[\langle E\rangle^*\ \delta E\big] + |\delta E|^2. \tag{11}$$

Averaging removes the cross term, since $\langle E\rangle$ is common to every realization while $\langle \delta E\rangle = 0$, leaving

$$\left\langle |E|^2 \right\rangle = |\,\langle E\rangle\,|^2 + \left\langle |\delta E|^2 \right\rangle \equiv I_{\mathrm{tot}}. \tag{12}$$

$I_{\mathrm{coh}} \equiv |\,\langle E\rangle\,|^2$ is the *coherent* intensity, i.e. the pattern produced by the mean mask, which we show below to be the three desired foci. $I_{\mathrm{halo}} \equiv \langle |\delta E|^2\rangle$ is the *halo* mentioned in Sec. 2: in a single realization the mask fluctuations produce a granular speckle pattern, and $I_{\mathrm{halo}}$ is its average over realizations. With many patches under the beam, the speckle of a single realization, averaged over its grains, already follows this envelope. Note that Eq. (12) holds for the average pattern: in any single realization the cross term survives, and it is what makes one *seed* differ from the ensemble. Here a seed is one random draw of the patch assignment. The two terms are evaluated in turn below, the remaining algebra being given in Secs. 2–4 of the supplementary material.

### 4.2 Coherent foci

Averaging Eq. (3) over realizations replaces each indicator by its probability,

$$\langle M(X,Y)\rangle = \sum_c \xi_c\, e^{i\phi_c(X,Y)}, \tag{13}$$

so the mean mask is a weighted superposition of the three ramps. The field to be Fraunhofer-transformed is $A_{\mathrm{SLM}}\,\langle M\rangle$, and each term is a Gaussian envelope carrying a linear phase, which Eq. (1) maps onto a Gaussian focus displaced by the ramp slope and scaled in amplitude by $\xi_c$. Summing the three displaced Gaussian foci and taking $I_{\mathrm{coh}}(x,y)$ gives cross terms between different foci, but since the foci are well separated ($d \gg w_0$) these cross terms are negligible, leaving the coherent intensity as a sum of three Gaussians,

$$I_{\mathrm{coh}}(x,y) = I_0 \sum_c \xi_c^2\, \mathcal{G}(x - x_c,\, y - y_c), \tag{14}$$

with $\mathcal{G}(x,y) = e^{-2(x^2+y^2)/w_0^2}$ and $I_0 = 2P_{\mathrm{tot}}/(\pi w_0^2)$ the peak intensity a single diffraction-limited focus would carry if it received the full power. The peak intensity therefore scales as $\xi_c^2$, the relation used to set the patch fractions in Sec. 2.3 and confirmed by the foci and dump intensities of Fig. 2(a). The patch size is absent from Eq. (14): on average, the foci are set by the channel fractions alone.

### 4.3 Speckle background

The fluctuation $\delta M$, i.e. the random arrangement of patches and the phase jumps at their boundaries, produces the speckle, whose ensemble average is $I_{\text{halo}}$. What controls it is the two-point covariance of the mask,

$$\begin{aligned}\Gamma(X_1, Y_1, X_2, Y_2) &\equiv \langle \delta M(X_1, Y_1)\, \delta M^*(X_2, Y_2)\rangle \\ &= \langle M(X_1, Y_1)\, M^*(X_2, Y_2)\rangle - \langle M(X_1, Y_1)\rangle \langle M^*(X_2, Y_2)\rangle ,\end{aligned} \tag{15}$$

which quantifies how the fluctuations of the mask at two points of the aperture are correlated. The mask itself is correlated over the whole aperture: in the limit $\xi_A \to 1$, for instance, every patch carries the ramp $\phi_A$. This long-range correlation is entirely contained in the mean mask $\langle M\rangle$ and produces the coherent foci of Eq. (14). The covariance $\Gamma$ only measures what is left once the mean is removed, and the patch structure splits it into two cases.

If the two points fall in different patches, their channels are drawn independently. The average of the product is then the product of the averages, and $\Gamma$ vanishes. The covariance is therefore non-zero only when the two points belong to the same patch, and the probability of this being the case is denoted $\Pi_\ell$. Within a patch, both points carry the same ramp $\phi_c$ with probability $\xi_c$, hence

$$\langle M(X_1, Y_1)\, M^*(X_2, Y_2)\rangle_{\text{same}} = \sum_c \xi_c\, e^{i[\phi_c(X_1,Y_1)-\phi_c(X_2,Y_2)]}. \tag{16}$$

The product of the means, on the other hand, draws the channels of the two points independently,

$$\langle M(X_1, Y_1)\rangle \langle M^*(X_2, Y_2)\rangle = \sum_{c,c'} \xi_c\, \xi_{c'}\, e^{i[\phi_c(X_1,Y_1)-\phi_{c'}(X_2,Y_2)]}. \tag{17}$$

Its diagonal terms, $c = c'$, carry the same ramps as Eq. (16) but with weight $\xi_c^2$ instead of $\xi_c$. Subtracting them leaves $\xi_c(1 - \xi_c)$, the variance of the channel assignment, which vanishes for $\xi_c \to 0$ or 1: a mask carrying a single ramp has no discontinuities and produces no halo. Since the ramps are linear, the remaining phase difference depends only on the separation, $\phi_c(X_1, Y_1) - \phi_c(X_2, Y_2) = \phi_c(r_X, r_Y)$ with $r_X = X_1 - X_2$ and $r_Y = Y_1 - Y_2$, so that this part of the covariance loses its dependence on where the pair sits,

$$\Gamma(r_X, r_Y) = \Pi_\ell(r_X, r_Y) \sum_c \xi_c(1 - \xi_c)\, e^{i\phi_c(r_X,r_Y)}. \tag{18}$$

The off-diagonal terms of Eq. (17), $c \neq c'$, are the cross-channel terms. They pair two *different* ramps and therefore depend on the position of the pair in the aperture as well as on its separation. They give weak halos between the foci, which are treated in Secs. 3.4 and 4.3 of the supplementary material and left out of Eq. (18).

The probability $\Pi_\ell$ is set by how far a point may be displaced before it crosses a patch boundary. In one dimension, a pair separated by $r_X$ stays inside the same patch of side length $\ell$ unless the first point lies within $|r_X|$ of an edge, which happens with probability $|r_X|/\ell$. The two directions are independent, so $\Pi_\ell = \Lambda(r_X/\ell)\Lambda(r_Y/\ell)$, where $\Lambda$ is the triangle function, $\Lambda(u) = \max(1 - |u|, 0)$ [37]: it equals $1 - |r_X|/\ell$, and zero once $|r_X|$ exceeds $\ell$.

Equation (18) depends on the separation alone, and this is what makes the focal-plane calculation tractable. Propagating $\delta M$ through Eq. (1) and averaging leaves a fourfold integral over $(X_1, Y_1)$ and $(X_2, Y_2)$. We recast it in terms of the separation and of the midpoint of the pair, $X_m = (X_1 + X_2)/2$ and $Y_m = (Y_1 + Y_2)/2$. The Gaussian envelope, of beam radius $w_{\text{SLM}}$, then factorizes exactly,

$$\begin{aligned}&A_{\text{SLM}}(X_m + \tfrac{r_X}{2},\, Y_m + \tfrac{r_Y}{2})\, A^*_{\text{SLM}}(X_m - \tfrac{r_X}{2},\, Y_m - \tfrac{r_Y}{2}) \\ &\quad = |A_{\text{SLM}}(X_m, Y_m)|^2\, \exp\left[-\frac{r_X^2 + r_Y^2}{2w_{\text{SLM}}^2}\right],\end{aligned} \tag{19}$$

so the midpoint integral acts on $|A_{\mathrm{SLM}}|^2$ alone and returns the total power $P_{\mathrm{tot}}$, leaving a two-dimensional transform over $(r_X, r_Y)$ of the patch correlation multiplied by the separation-dependent Gaussian. The factor $e^{i\phi_c}$ then displaces this transform, by the same mechanism that displaced the coherent foci in Eq. (14): a linear phase in the aperture becomes a displacement in the focal plane, and the two directions separate. The resulting one-dimensional patch kernel is

$$\mathcal{S}(u) = \int_{-\infty}^{\infty} \Lambda(r/\ell)\, \exp\left[-\frac{r^2}{2w_{\mathrm{SLM}}^2}\right] \exp\left[-i\frac{k_0\, u\, r}{f}\right] dr, \qquad k_0 = \frac{2\pi}{\lambda}, \tag{20}$$

which, for $\ell \ll w_{\mathrm{SLM}}$, reduces to the diffraction pattern of a single patch of width $\ell$ (Sec. 4.2 of the supplementary material). Each channel therefore carries its own halo, centered on its own focus, the product of two such kernels, reading

$$I_{\mathrm{halo}}(x, y) = \frac{P_{\mathrm{tot}}}{(\lambda f)^2} \sum_c \xi_c(1 - \xi_c)\, \mathcal{S}(x - x_c)\, \mathcal{S}(y - y_c). \tag{21}$$

Small patches diffract strongly and spread the halo thinly over the whole focal field. Large patches diffract weakly and concentrate the halo around each focus. In the extreme case where only one patch covers the beam, a single realization is one ramp drawn at random, and there is no halo to be spread.

The randomness of the assignment is what makes this background featureless. A periodic partition, such as a checkerboard of patches, would carry the same power $1 - \sum_c \xi_c^2$ outside the main foci, but would concentrate it into discrete grating orders around each focus, on the scale $\lambda f/\ell$, i.e. into secondary foci.

# 5 Validation

$I_{\mathrm{tot}}(x, y)$ is an ensemble average, and a single mask produces a focus whose intensity fluctuates around that average. To compare with this ensemble average, we experimentally generate $N = 100$ independent random masks per patch size and record the focal intensity for each. Figure 3 shows the cut $y = 0$ for two patch sizes $s$, overlaid on the same axes, with $\alpha = 1/2$ and no dump so that only channels $A$ and $B$ are populated.

The figure compares three quantities for each $s$. The red curve is the experiment, that is the average of the $N = 100$ measured camera frames. The orange curve is the simulation: a direct Fraunhofer transform of $N = 100$ random masks, averaged. The dashed black curve is $I_{\mathrm{tot}}(x, y)$ evaluated for the same parameters.

For both patch sizes, intensities are expressed in decibels relative to the coherent peak $I_{\mathrm{peak}}$, the maximum of $I_{\mathrm{coh}}(x, y)$ (Eq. (14)). The decibel scale is defined as $10 \log_{10}(I/I_{\mathrm{peak}})$, so that 0 dB marks that peak and negative values are weaker than it. Two contributions limit what the camera can see, and both are properties of the setup rather than of the mask. Far from the foci the noise floor comes from the detector itself, near $-35$ to $-40$ dB, set by the camera readout and ambient stray light. Closer to the foci, the contribution to the background originates from the beam itself: even with a flat mask, a pedestal of scattered light appears around $-20$ dB, from imperfect beam quality and scattering in the optics upstream (Fig. S4 of the supplementary material).

The coherent foci are recovered in every case, and the halo stays at least 20 dB below. What changes with $s$ is the halo alone, i.e. the speckle background $I_{\mathrm{halo}}(x, y)$ produced by the mask fluctuations, whose dependence on $s$ enters only through the patch kernel $\mathcal{S}$ of Eq. (20). As $s$ grows from 8 to 64 the halo narrows in space and rises in amplitude, gaining visible structure as it emerges from the noise floor. The total power carried by $I_{\mathrm{halo}}(x, y)$ (Eq. (21)) is unchanged, but is redistributed from across the whole field into the immediate neighborhood of the foci, so its level there rises from about $-38$ dB to about $-22$ dB across the range.

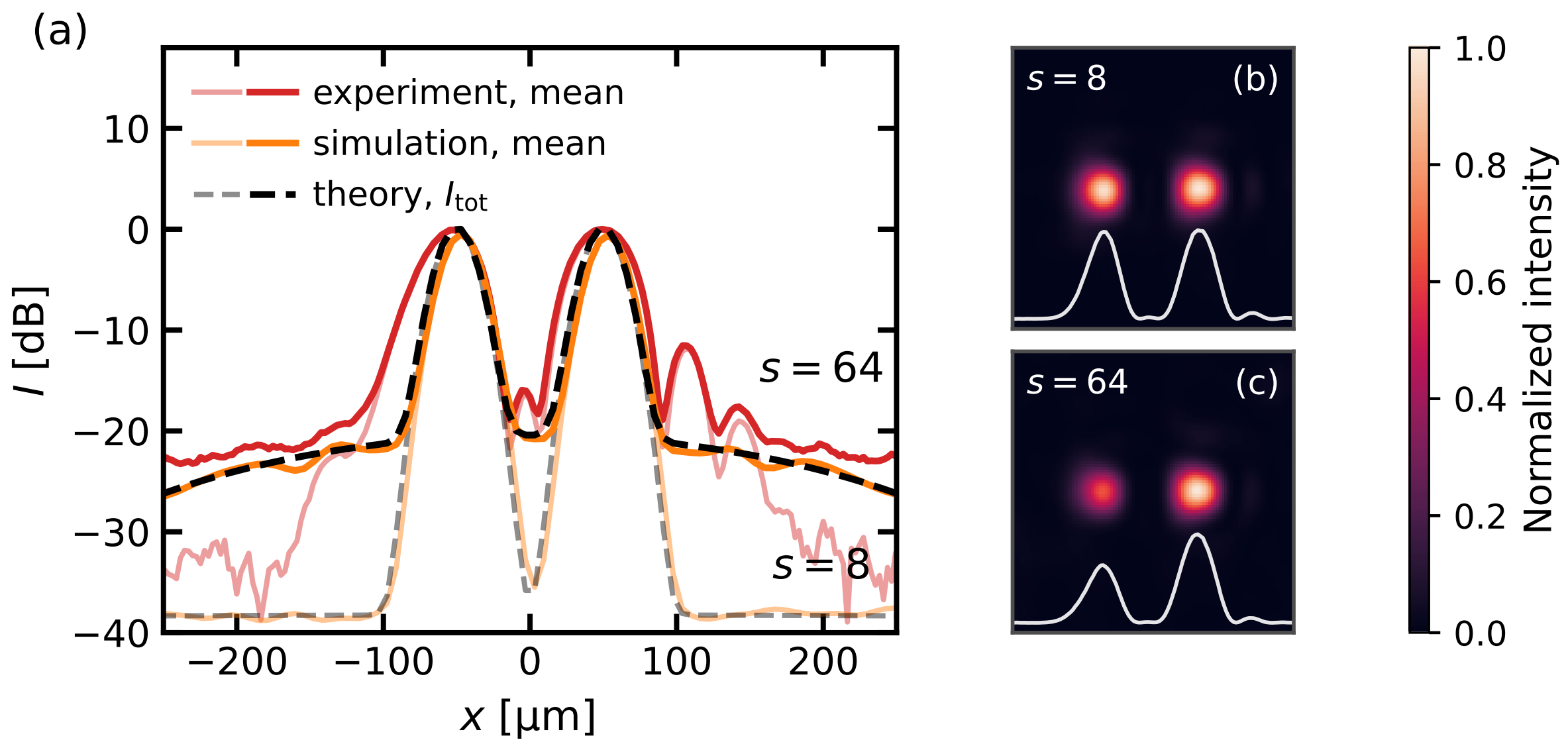

**Figure 3:** (a) Validation of $I_{\text{tot}}(x, y)$, shown as focal-plane intensity cuts at $y = 0$ on a log scale, for two patch sizes $s = 8, 64$ ($\alpha = 1/2$, no dump) overlaid on the same axes: faint curves and the $s = 8$ label for the smaller patch, bold curves and the $s = 64$ label for the larger one. Each curve overlays the measured average over $N = 100$ camera frames, a direct numerical Fraunhofer propagation of $N = 100$ random masks, and the prediction of $I_{\text{tot}}(x, y)$ (dashed). Intensities are normalized so that the theory peak sits at 0 dB. The measurement is bounded from below by the camera noise floor near −35 to −40 dB and, around the foci, by the intrinsic pedestal of the laser beam near −20 dB. The halo stays below this floor at $s = 8$ and rises into visible lobes at $s = 64$: the measured tail is markedly higher for the larger patch, in accordance with theory, while the coherent foci are recovered in both cases. (b), (c) Corresponding single-seed beam images at $s = 8$ and $s = 64$, with the same $y = 0$ cut traced in white on a linear scale.

Figures 3(b) and (c) tell the complementary story about the foci themselves. At $s = 8$ the pair of foci looks clean, while at $s = 64$ the spots visibly degrade in shape and intensity balance. The seed-to-seed fluctuations, small near the foci and large in the halo, grow with $s$ as fewer independent patches cover the beam. Agreement on the halo envelope does not by itself make the largest patch sizes usable, since the focus itself becomes seed-dependent as $s$ grows, an effect that the logarithmic scale of the cuts in Fig. 3(a) hides but panels (b), (c) already suggest, and which is taken up in the following section.

## 6 Choosing the patch size

The patch size $s$ is the one parameter of the method left free. It is absent from $I_{\text{coh}}(x, y)$ (Eq. (14)), present only through the kernel $\mathcal{S}$ in $I_{\text{halo}}(x, y)$ (Eq. (21)), so in the ensemble average the foci are set by the channel fractions alone. A single displayed mask is not the ensemble, however, and $s$ impacts the measured foci, in two ways: the halo moves in toward them as $s$ grows, and the realized channel fractions drift further from the values targeted. Both effects stay far below the coherent peak over most of the accessible range, and are further suppressed in the nonlinear applications that motivate this work, whose response scales as a high power of the local intensity. In practice, the patch size $s$ still has to be chosen, and theory and experiment do not point the same direction.

### 6.1 Arguments for small patches

The first argument is the halo shape. A small patch diffracts strongly and scatters the light far from the coherent peaks. A large patch diffracts weakly, so the same power is deposited close to the foci, and may contaminate them. Figure 3 shows this directly: the halo stays below the bottom of the panel at $s = 8$ and rises to about $-22$ dB at $s = 64$. Growing the patch does not create background light, it only moves the existing background closer to the foci.

The second argument is statistical. $I_{\text{tot}}(x, y)$ is an ensemble average, but in practice a single patch assignment is displayed at a time. The beam does the averaging instead: it illuminates many patches at once, and what reaches the foci is the average over them. The more patches fit across the beam, the closer the realized channel fractions sit to their targets. Once the patch approaches the beam size, only a handful are lit, the averaging fails, and the realized intensities depend on the particular draw. This is the seed-to-seed variation: narrow at small $s$, where many patches are averaged, and wide at large $s$, where few are. It is also what Fig. 3(b), (c) show. At $s = 64$ the targeted $\alpha = 1/2$ is not realized: the draw that happens to be displayed sends more area to one channel than to the other, and the two foci come out unequal.

### 6.2 Fringing fields and the residual zeroth order

The model assumes the SLM reproduces the targeted phase exactly. However, this is not entirely true. The liquid-crystal orientation follows the smooth field set by the pixel voltages, and elastic forces in the layer resist abrupt changes between neighboring pixels [39–42]. These *fringing-field* effects act as a spatial low-pass filter on the displayed phase. Every boundary between patches of different channels carries a large phase jump, and the filter turns those boundaries into strips where the targeted ramp is not reproduced. A linear phase ramp acts like a diffraction grating, deflecting light into the intended, or *first*, order. Light impinging on the boundary strips may not be deflected in the intended direction, in which case it does not participate in the coherent foci. Part of it stays in the undeflected *zeroth* order, the rest is scattered to larger angles.

### 6.3 Operating window

The two mechanisms from Sec. 6.1 and Sec. 6.2, seed-to-seed variation and fringing loss, bound the patch size from both sides. Below $s \simeq 4$, fringing loss dominates: a significant portion of the light stays in the zeroth order [Fig. 4(a)], and correspondingly little reaches the first order [Fig. 4(b)]. Above $s \simeq 16$, seed-to-seed variation dominates: too few patches cover the beam, and the realized intensity ratio depends on the seed rather than on the targeted parameters. Both limits are qualitative, marking where each mechanism becomes visible. The targeted parameters $\alpha$, $\beta_A$, $\beta_B$ are continuous: they set assignment probabilities rather than displayed phase levels, and how closely a given draw realizes them is limited by the number of patches under the beam, i.e. by the seed-to-seed spread above.

## 7 Conclusion

A randomly partitioned phase mask turns a single SLM into a common-path beam splitter whose beam intensities are set by patch fractions alone. The mask follows in closed form from the desired fractions and ramp slopes, with no iteration, and a single random draw produces it. We measured continuous tuning of the intensity ratio from 0 to 1 with both spatial modes preserved and the relative phase constant to a few tens of mrad, and we showed that the focal intensity separates into the desired foci and a halo whose shape is set by the patch size alone.

The cost of the random encoding is the halo, which carries a fraction $1 - \sum_c \xi_c^2$ of the total power $P_{\text{tot}}$. That fraction is not a defect of the implementation but a property of any mask that multiplexes several ramps on one aperture. Deterministic gratings perform better at fixed ratios: the optimal triplicator sends most of the power into its three beams [16, 17], but the ratios are then written into the profile once and for all. What the random partition enables is the continuous, independent intensity control that a fixed geometry cannot provide. Fixing the random seed fixes the mask exactly, so nothing is lost in reproducibility. What a given seed delivers is the response of one member of the ensemble rather than of the ensemble itself. Once chosen and characterized, a seed can simply be kept.

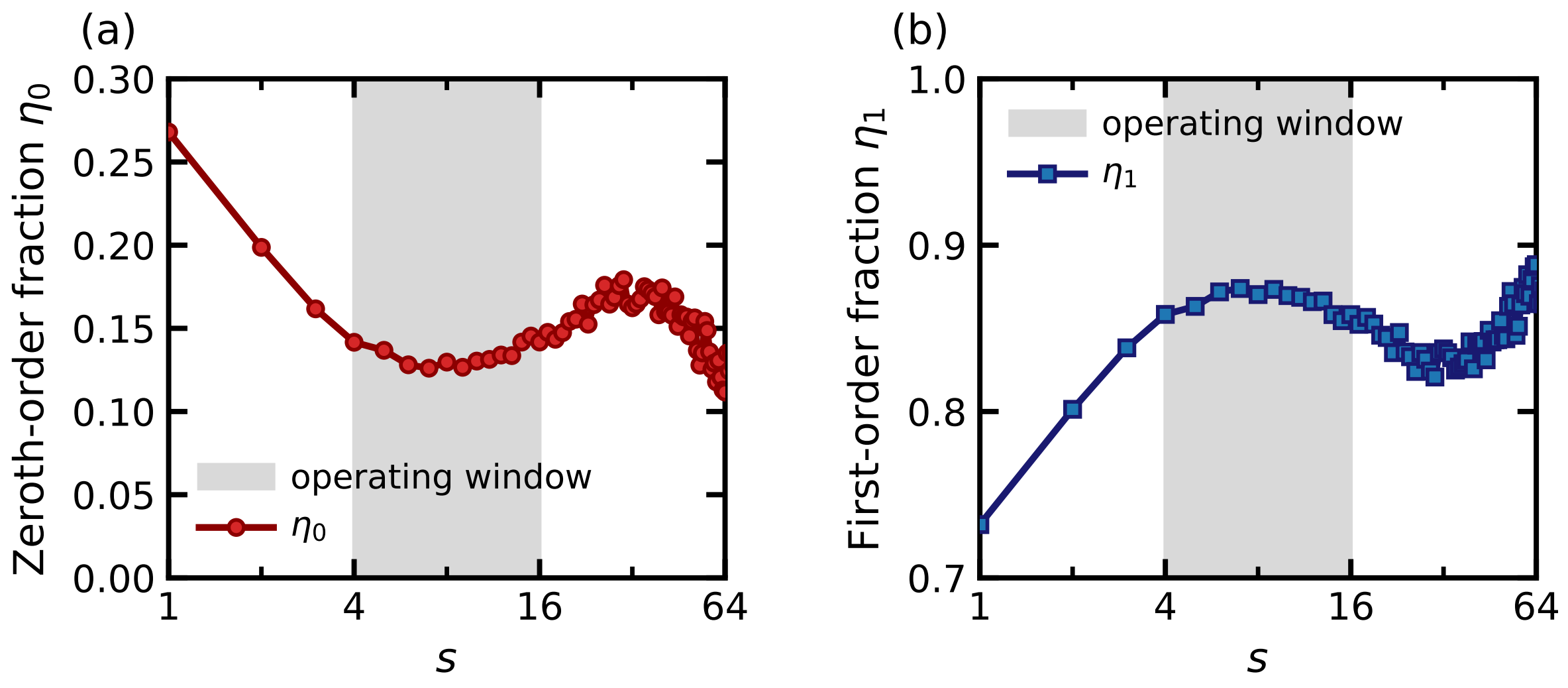

**Figure 4:** Split of the measured power between the zeroth and first orders as a function of patch size $s$, with (a) $\eta_0 = P^{(0)}/(P^{(0)} + P^{(1)})$ and (b) $\eta_1 = P^{(1)}/(P^{(0)} + P^{(1)})$. Fringing-field effects scatter light out of the first order, and $\eta_1$ rises as the boundary density falls. The shaded band marks the operating window $s \simeq 4$–16, bounded below by this loss and above by the seed-to-seed spread of Sec. 6.1, which appears here as the growing scatter of the points at large $s$.

The patch size is the one parameter left to choose, and it is bounded from both sides for unrelated reasons: small patches raise the fringing loss, which follows the inter-patch boundary density, while large patches leave too few patches under the beam and make the realized ratio seed-dependent. The optimum is therefore set by the device rather than by the theory. The fringing loss is not necessarily fixed either: since it follows the total inter-patch boundary length, patch layouts designed to minimize that length, rather than a plain square grid, may reduce it further.

This sensitivity to fringing fields is also why amplitude encoding was not pursued as an alternative to the random phase partition. A phase-only SLM can encode a prescribed amplitude by locally reducing the depth of a blazed carrier grating, or by related complex-amplitude encoding schemes [43–45], which could in principle attenuate one half of a $\pi$-step mask. These encodings, however, tie the amplitude control to the fidelity of the displayed phase: the attenuation acts through the local modulation depth, which Sec. 6.2 shows to be distorted by fringing fields, so that amplitude and phase errors of the beams become coupled.

The method suits applications needing a few phase-locked channels with tunable intensities, rather than arbitrary complex-amplitude shaping. It sits between fixed-geometry elements such as phase plates and birefringent wedges, which offer no dynamic intensity control, and iterative computer-generated holography [32–36], which reaches arbitrary targets at the cost of optimization loops, careful calibration, and masks without interpretable structure.

Three extensions follow directly. (i) The partition generalizes to $n$ channels with arbitrary ramps. (ii) Replacing a ramp by an arbitrary mode-defining phase, and removing the separating tilt, places several spatial modes at one focus with the patch fractions setting their relative weights, which turns the splitter into a mode multiplexer with continuously tunable mode content. (iii) And since the lower bound on $s$ is instrumental, model-based pre-distortion of the displayed pattern [46–48] should decrease it and widen the useful range. The characterization reported here is in linear optics, and applying the splitter to the phase characterization of high-order harmonic generation is the subject of forthcoming work.

## Funding

The authors acknowledge support from the Swedish Research Council (Grant Nos. 2021-04691, 2022-03519) and the Crafoord Foundation. This project was also financially supported by the Swedish Foundation for Strategic Research (FFL24-0144). A.-L.V. acknowledges support from Åforsk through the Swedish Foundations' Starting Grant (RAVIOLI).

This work was also supported by state funding managed by the National Research Agency under the France 2030 investment plan bearing the reference ANR-22-EXLU-0002 and the moonshot project "Tornado"; it received funding from the Agence Nationale pour la Recherche (contract No. ANR HELIMAG ANR-21-CE30-0037) and the Indo-French CEFIPRA Grant Project No. 7104-I. It was funded by the EU's Horizon Europe programme under grant agreement number 101131771 Lasers4EU [project 75540]. Views and opinions expressed are however those of the author(s) only and do not necessarily reflect those of the European Union. The European Union cannot be held responsible for them.

## Supplementary material

The supplementary material gives the power balance of the $\beta_A$ scan including the dump, the full derivation of $I_{\text{tot}}(x, y)$, including the cross-channel terms omitted in Eq. (18), and the focal-plane intensity of the unshaped beam.

# Supplementary material: Phase-locked beam splitting with continuously tunable intensity ratio using stochastic phase masks

Melvin Redon[1,*], Titouan Gadeyne[2], Gaspard Beaufort[1], Saga Westerberg[1], Vartika Vishnoi[2], Peter Smorenburg[1,3], Anne-Lise Viotti[1], Thierry Ruchon[2] and Cord L. Arnold[1]

[1]Department of Physics, Lund University, P.O. Box 118, SE-22100 Lund, Sweden
[2]Université Paris-Saclay, CEA, LIDYL, 91191 Gif-sur-Yvette, France
[3]ASML Research, ASML Netherlands B.V., 5504 DR Veldhoven, The Netherlands
[*]Author to whom any correspondence should be addressed.
E-mail: melvin.redon@fysik.lu.se

**Conventions.** Notation follows the main text (Sec. 2): the ramps $\phi_c$, the mask $M$ and the focus positions $(x_c, y_c)$ are defined there, where $c \in \{A, B, D\}$ labels the channel. The focal field is calculated within the Fraunhofer approximation

$$E(x,y) \propto \frac{1}{i\lambda f} \iint E_{\mathrm{SLM}}(X,Y)\, e^{-i\frac{2\pi}{\lambda f}(Xx+Yy)}\, dX\, dY, \tag{S1}$$

with $\iint |E|^2\, dx\, dy = P_{\mathrm{tot}}$. Splitting the mask into its mean and a fluctuation, $M = \langle M\rangle + \delta M$ with $\langle \delta M\rangle = 0$, carries through to the focal field, $E = \langle E\rangle + \delta E$, and the ensemble-averaged focal intensity splits accordingly into a coherent part and a halo part,

$$I_{\mathrm{tot}}(x,y) \equiv \left\langle |E|^2 \right\rangle = \underbrace{|\langle E\rangle|^2}_{\equiv\, I_{\mathrm{coh}}} + \underbrace{\left\langle |\delta E|^2 \right\rangle}_{\equiv\, I_{\mathrm{halo}}}. \tag{S2}$$

Section 1 first gives the power balance of the $\beta_A$ scan of Fig. 2 of the main text. The rest of this document derives $I_{\mathrm{coh}}$ (Sec. 2) and $I_{\mathrm{halo}}$ (Secs. 3-4) in full, keeping the cross-channel terms dropped in the main text, and recombines them into $I_{\mathrm{tot}}$ in Sec. 4.4.

## 1 Power in the dump

Figure 2(a) of the main text shows each focus normalized separately. Figure S1 shows the same $\beta_A$ scan in counts: the power collected in the two main foci, $I_A + I_B$, and with the dump added, $I_A + I_B + I_D$. With $\alpha = 1/2$ and $\beta_B = 0$, Eq. (7) of the main text gives $\xi_A = \sqrt{1-\beta_A}/2$, $\xi_B = 1/2$ and $\xi_D = (1 - \sqrt{1-\beta_A})/2$. Since $P_c = \xi_c^2\, P_{\mathrm{tot}}$,

$$I_A + I_B \propto 2 - \beta_A, \qquad I_A + I_B + I_D \propto 2 - \beta_A + \left(1 - \sqrt{1-\beta_A}\right)^2. \tag{S3}$$

The power removed from $A$ is not all collected by the dump. The sum over the three foci drops to 3/4 of its initial value at $\beta_A = 3/4$, where $A$ and $D$ share the $A$-side patches equally, and returns to its initial value at $\beta_A = 1$, where all of them feed the dump. The missing power is diffracted into the halo [mechanism (ii) of Sec. 2.3 of the main text]. The measurement follows this prediction across the scan, with a minimum of 0.74 at $\beta_A = 0.75$.

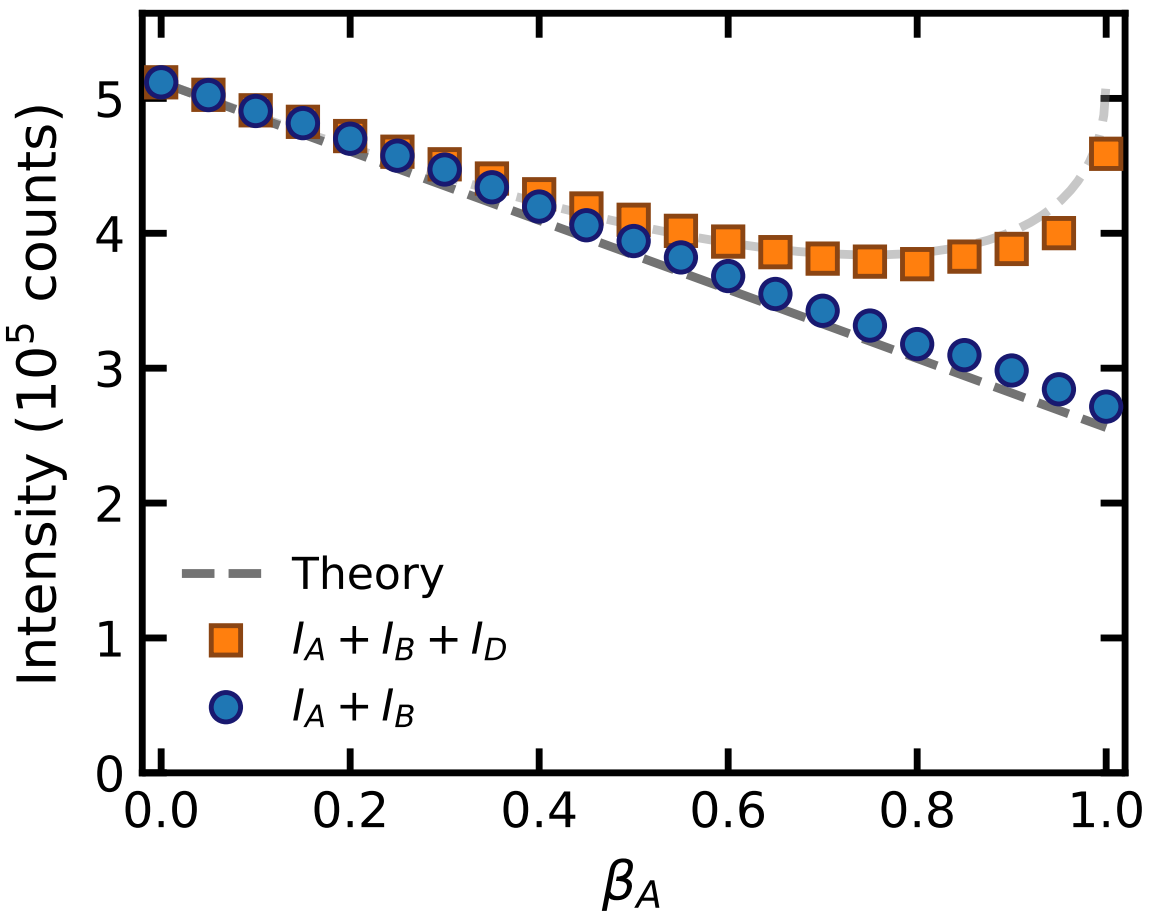

**Figure S1:** Power in the two main foci, $I_A + I_B$, and including the dump, $I_A + I_B + I_D$, against $\beta_A$ ($\alpha = 1/2$, $\beta_B = 0$). Dashed lines: mask model, scaled to the $\beta_A = 0$ point.

# 2 Mean field and coherent foci

## 2.1 Mean mask

Patches are independent draws, so the ensemble mean of the mask replaces each indicator by its probability,

$$\langle M(X,Y)\rangle = \sum_c \xi_c \, e^{i\phi_c(X,Y)}, \tag{S4}$$

where $\phi_c$ is the linear phase ramp of Eq. (2) of the main text, i.e. the phase of a wedge. The SLM is pixelated, so it displays a staircase version of each ramp, constant within each pixel, and therefore acts as a grating that sends a small fraction of the light into higher diffraction orders. This correction is negligible for the ramps used here, and we drop it throughout.

## 2.2 Mean field at the foci

Inserting the mean mask of Eq. (S4) in Eq. (S1), and evaluating the Gaussian integral

$$\int e^{-X^2/w_{\rm SLM}^2} e^{-ikX}\, dX = \sqrt{\pi}\, w_{\rm SLM}\, e^{-w_{\rm SLM}^2 k^2/4}$$

gives, for channel $A$,

$$\langle E\rangle_A = \frac{\xi_A\, A_0\, \pi w_{\rm SLM}^2}{i\lambda f}\, \exp\left[-\frac{\pi^2 w_{\rm SLM}^2}{\lambda^2 f^2}\Big((x-x_A)^2 + y^2\Big)\right]. \tag{S5}$$

The three foci are well separated ($d \gg w_0$), so cross terms between them are negligible,

$$|\langle E\rangle|^2 \approx \frac{\pi^2 w_{\rm SLM}^4\, A_0^2}{(\lambda f)^2} \sum_c \xi_c^2\, \mathcal{G}(x - x_c,\, y - y_c), \qquad \mathcal{G}(x,y) = e^{-2(x^2+y^2)/w_0^2}. \tag{S6}$$

The prefactor is fixed by the total power, $\iint |A_{\rm SLM}|^2\, dX\, dY = A_0^2\, \pi w_{\rm SLM}^2/2 = P_{\rm tot}$, so that, using $w_0 = \lambda f/(\pi w_{\rm SLM})$,

$$\frac{\pi^2 w_{\rm SLM}^4\, A_0^2}{(\lambda f)^2} = \frac{2\pi w_{\rm SLM}^2 P_{\rm tot}}{(\lambda f)^2} = \frac{2P_{\rm tot}}{\pi w_0^2} \equiv I_0, \tag{S7}$$

and Eq. (S6) becomes

$$I_{\rm coh}(x,y) \equiv |\langle E\rangle|^2 = I_0 \sum_c \xi_c^2\, \mathcal{G}(x - x_c,\, y - y_c). \tag{S8}$$

Here $I_0 = 2P_{\rm tot}/(\pi w_0^2)$ is the peak intensity of a single diffraction-limited focus carrying the full power.

# 3 Two-point covariance of the mask fluctuation

## 3.1 Decomposition and the covariance to be computed

$I_{\rm coh}$, computed above in Eq. (S8), is the first term of the decomposition of Eq. (S2). This section computes the covariance needed for the second term, $I_{\rm halo} \equiv \langle |\delta E|^2 \rangle$, the speckle background. The decomposition follows the standard treatment of speckle from a random screen [1, 2], adapted from the binary amplitude screen to our multi-channel phase partition.

Because $\delta E$ obeys Eq. (S1) with $M$ replaced by $\delta M$, squaring and averaging gives

$$\left\langle |\delta E(x,y)|^2 \right\rangle = \frac{1}{(\lambda f)^2} \iint\!\!\iint A_{\rm SLM}(X_1,Y_1)\, A^*_{\rm SLM}(X_2,Y_2)\, \Gamma(X_1,Y_1,X_2,Y_2) \times e^{-i\frac{2\pi}{\lambda f}[(X_1-X_2)x+(Y_1-Y_2)y]}\, dX_1\, dY_1\, dX_2\, dY_2, \quad \text{(S9)}$$

where the only stochastic quantity is the two-point covariance of the mask fluctuation,

$$\begin{aligned}\Gamma(X_1,Y_1,X_2,Y_2) &\equiv \langle \delta M(X_1,Y_1)\, \delta M^*(X_2,Y_2) \rangle \\ &= \langle M(X_1,Y_1)\, M^*(X_2,Y_2) \rangle - \langle M(X_1,Y_1) \rangle \, \langle M^*(X_2,Y_2) \rangle \, . \end{aligned} \quad \text{(S10)}$$

Two points in different patches have independent channel assignments, so the average of their product factorizes and the covariance vanishes there. Two points in the same patch share the same assignment and are correlated. Writing $M_1 \equiv M(X_1,Y_1)$ and $M_2 \equiv M(X_2,Y_2)$ for short and conditioning on the same-patch event,

$$\langle M_1 M_2^* \rangle = \Pi_\ell \, \langle M_1 M_2^* \rangle_{\rm same} + (1-\Pi_\ell)\, \langle M_1 \rangle \, \langle M_2^* \rangle \, , \quad \text{(S11)}$$

so the different-patch piece cancels against the subtraction in Eq. (S10) and leaves

$$\Gamma = \Pi_\ell(r_X, r_Y) \left[ \langle M_1 M_2^* \rangle_{\rm same} - \langle M_1 \rangle \, \langle M_2^* \rangle \right], \quad \text{(S12)}$$

where $\Pi_\ell(r_X, r_Y)$ is the probability that two points separated by $(r_X, r_Y)$ lie in the same patch.

## 3.2 Same-patch probability

Since the covariance is an ensemble average, $\Pi_\ell$ can be computed by averaging over where the pair sits within a patch, using translational symmetry to fix the patch itself. Place $X_1$ uniformly in a patch of side length $\ell$ along $X$, and ask whether the second point, $X_2 = X_1 - r_X$, falls in the same patch. It does unless $X_1$ lies within a distance $|r_X|$ of the patch edge that $r_X$ points toward, since only then does subtracting $r_X$ push $X_2$ past that edge. Because $X_1$ is uniform over a patch of width $\ell$, the probability of landing in that edge strip is $|r_X|/\ell$, so the probability of staying in the same patch is $1 - |r_X|/\ell$ for $|r_X| < \ell$, and zero once $|r_X|$ exceeds $\ell$ (Fig. S2). Independence in $X$ and $Y$ gives

$$\Pi_\ell(r_X, r_Y) = \Lambda\!\left(\tfrac{r_X}{\ell}\right) \Lambda\!\left(\tfrac{r_Y}{\ell}\right), \qquad \Lambda(u) = \max(1-|u|, 0), \quad \text{(S13)}$$

with $\Lambda$ the triangle function. This expression is an average over the position of the pair within a patch.

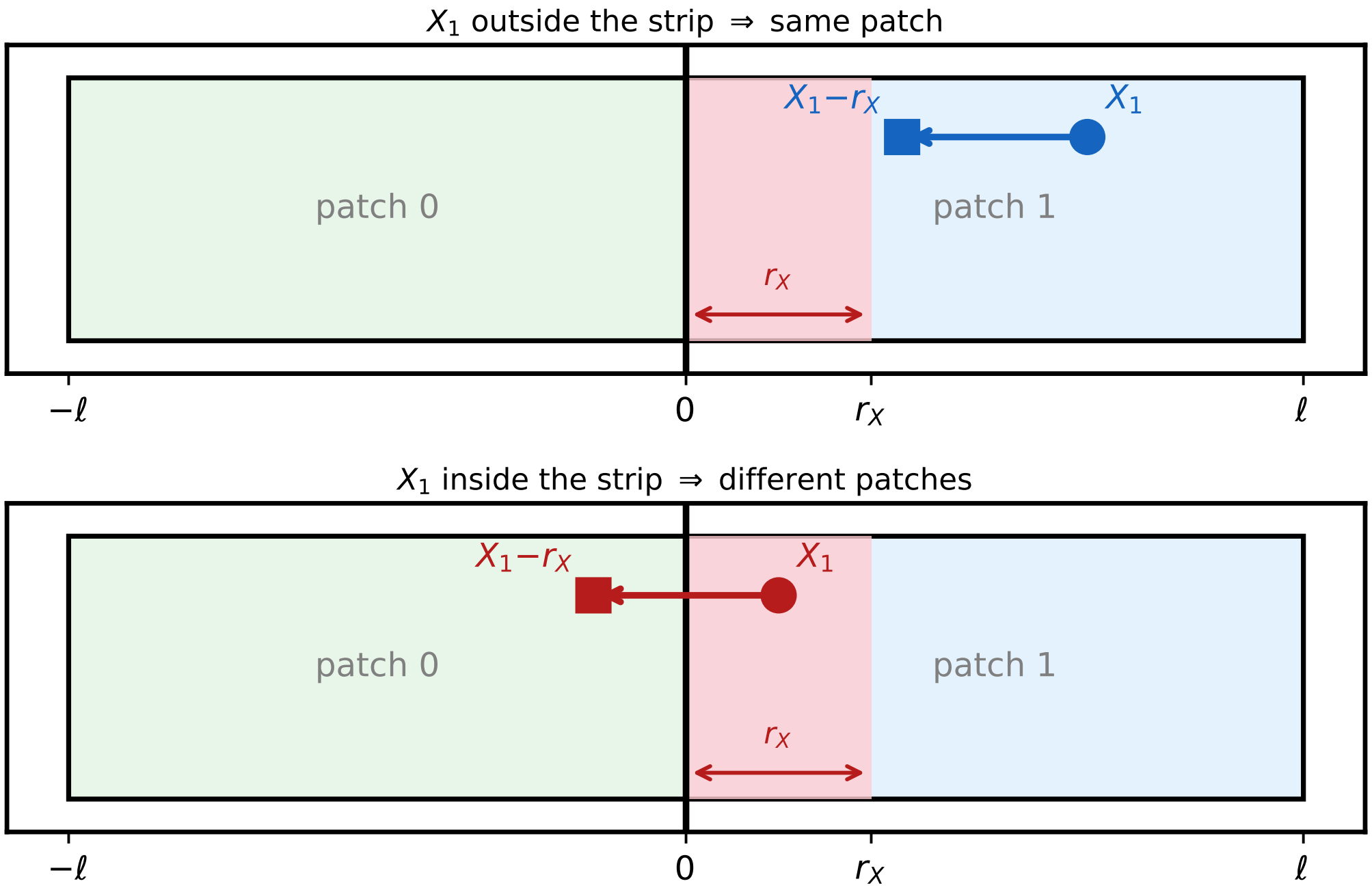

**Figure S2:** Geometric derivation of the same-patch probability in one dimension. The strip of width $|r_X|$ (red) marks the positions of $X_1$ for which $X_1 - r_X$ crosses into the adjacent patch. The probability of sharing a patch is the fraction of the patch outside the strip, $1 - |r_X|/\ell$.

### 3.3 Same-patch contribution

When two points share a patch they are assigned to the same channel $c$ with probability $\xi_c$,

$$\langle M_1 M_2^* \rangle_{\text{same}} = \sum_c \xi_c \, e^{i[\phi_c(X_1,Y_1) - \phi_c(X_2,Y_2)]}. \tag{S14}$$

The ramps are linear, so their difference depends only on the separation, and

$$\langle M_1 M_2^* \rangle_{\text{same}} = \sum_c \xi_c \, e^{i\phi_c(r_X, r_Y)}. \tag{S15}$$

### 3.4 Different-patch contribution

From Eq. (S4), the product of the means is

$$\langle M_1 \rangle \langle M_2^* \rangle = \sum_{c,c'} \xi_c \xi_{c'} \, e^{i[\phi_c(X_1,Y_1) - \phi_{c'}(X_2,Y_2)]}. \tag{S16}$$

The diagonal $c = c'$ has the same form as Eq. (S15) but with weight $\xi_c^2$ instead of $\xi_c$, so subtracting it leaves the variance $\xi_c(1 - \xi_c)$ of the channel assignment. The off-diagonal $c \neq c'$ terms are the cross-channel piece. They do not reduce to a function of the separation, and it is convenient to introduce the midpoint of the pair,

$$X_m = \tfrac{X_1 + X_2}{2}, \quad Y_m = \tfrac{Y_1 + Y_2}{2}, \quad r_X = X_1 - X_2, \quad r_Y = Y_1 - Y_2. \tag{S17}$$

Writing $X_j = X_m \pm r_X/2$ and likewise for $Y_j$, the phase of an ordered term $(c, c')$ splits into a midpoint part and a separation part. Since $\Gamma(2,1) = \Gamma^*(1,2)$, the two orderings of a pair must be summed together, and they share the same separation factor while carrying opposite midpoint phases. Each unordered pair therefore contributes a single midpoint cosine times a single separation exponential. With three channels there are three such pairs. For $\{A, B\}$ both ramps lie along $X$, so the separation phases cancel and only a midpoint term survives,

$$C_{AB}(X_m) = 2\,\xi_A \xi_B\,\cos(2k_t X_m). \tag{S18}$$

For the two mixed pairs the ramps lie along different axes, and both a midpoint cosine and a separation phase remain,

$$C_{AD} = 2\,\xi_A \xi_D\,\cos\!\big(k_t X_m - n_D k_t Y_m\big)\,e^{+i(k_t r_X + n_D k_t r_Y)/2}, \tag{S19}$$

$$C_{BD} = 2\,\xi_B \xi_D\,\cos\!\big(k_t X_m + n_D k_t Y_m\big)\,e^{-i(k_t r_X - n_D k_t r_Y)/2}. \tag{S20}$$

### 3.5 Full covariance

Combining Eqs. (S12), (S15) and the diagonal of the different-patch term,

$$\boxed{\begin{aligned}\Gamma(X_m, Y_m, r_X, r_Y) = \Pi_\ell(r_X, r_Y)\Big[\sum_c \xi_c(1-\xi_c)\,e^{i\phi_c(r_X, r_Y)} \\ - C_{AB} - C_{AD} - C_{BD}\Big].\end{aligned}} \tag{S21}$$

The first term depends on the separation alone and produces the halo around each focus, with weight $\xi_c(1-\xi_c)$. It is the only term kept in the main text. The cross-channel piece carries midpoint frequencies $2k_t$ for $\{A, B\}$ and $k_t\sqrt{1+n_D^2}$ for the mixed pairs.

## 4 Mean focal-plane intensity

### 4.1 Setting up the integral

Changing Eq. (S9) to midpoint and separation coordinates, the Gaussian envelope factorizes exactly,

$$\begin{aligned} &A_{\mathrm{SLM}}\big(X_m + \tfrac{r_X}{2}, Y_m + \tfrac{r_Y}{2}\big)\,A^*_{\mathrm{SLM}}\big(X_m - \tfrac{r_X}{2}, Y_m - \tfrac{r_Y}{2}\big) \\ &\quad = |A_{\mathrm{SLM}}(X_m, Y_m)|^2\,e^{-(r_X^2 + r_Y^2)/(2w_{\mathrm{SLM}}^2)}. \end{aligned} \tag{S22}$$

The speckle term becomes

$$\begin{aligned} I_{\mathrm{halo}}(x, y) \equiv \left\langle |\delta E|^2 \right\rangle = \frac{1}{(\lambda f)^2} \iiiint |A_{\mathrm{SLM}}(X_m, Y_m)|^2\,\Gamma \\ \times\, e^{-(r_X^2 + r_Y^2)/(2w_{\mathrm{SLM}}^2)}\,e^{-i(k_x r_X + k_y r_Y)}\,dX_m\,dY_m\,dr_X\,dr_Y, \end{aligned} \tag{S23}$$

with focal-plane wavenumbers $k_x = k_0 x/f$ and $k_y = k_0 y/f$. Substituting Eq. (S21) splits $I_{\mathrm{halo}}$ into a separation-only part and a cross-channel part, computed in turn below.

### 4.2 Halos at the foci

For the separation-only part the midpoint integral factors out and gives $\iint |A_{\mathrm{SLM}}|^2\,dX_m\,dY_m = P_{\mathrm{tot}}$. The remaining integral over $(r_X, r_Y)$ contains the ramp $e^{i\phi_c}$, which shifts the kernel onto the corresponding focus: for channel $A$ the ramp $\phi_A = k_t r_X$ takes $k_x$ to $k_x - k_t = k_0(x - x_A)/f$. Writing the kernel directly as a function of focal-plane position,

$$\mathcal{S}(u) = \int_{-\infty}^{\infty} \Lambda\!\left(\frac{r}{\ell}\right) e^{-r^2/(2w_{\mathrm{SLM}}^2)}\,e^{-ik_0 u r/f}\,dr, \tag{S24}$$

the two directions separate and the total is

$$\left\langle |\delta E|^2 \right\rangle_{\mathrm{halo}} = \frac{P_{\mathrm{tot}}}{(\lambda f)^2} \sum_c \xi_c(1-\xi_c)\,\mathcal{S}(x - x_c)\,\mathcal{S}(y - y_c). \tag{S25}$$

The kernel is the transform of a product of two windows on the separation, and the narrower of the two sets its width. In the two limits,

$$\mathcal{S}(u) \approx \begin{cases} \ell \,\mathrm{sinc}^2\!\left(\dfrac{u\,\ell}{\lambda f}\right) & \ell \ll w_{\mathrm{SLM}} \quad \text{(patch-limited)}, \\ \sqrt{2\pi}\, w_{\mathrm{SLM}}\, e^{-2u^2/w_0^2} & \ell \gg w_{\mathrm{SLM}} \quad \text{(beam-limited)}, \end{cases} \tag{S26}$$

as illustrated in Fig. S3. In the patch-limited regime the halo is the diffraction pattern of a single patch.

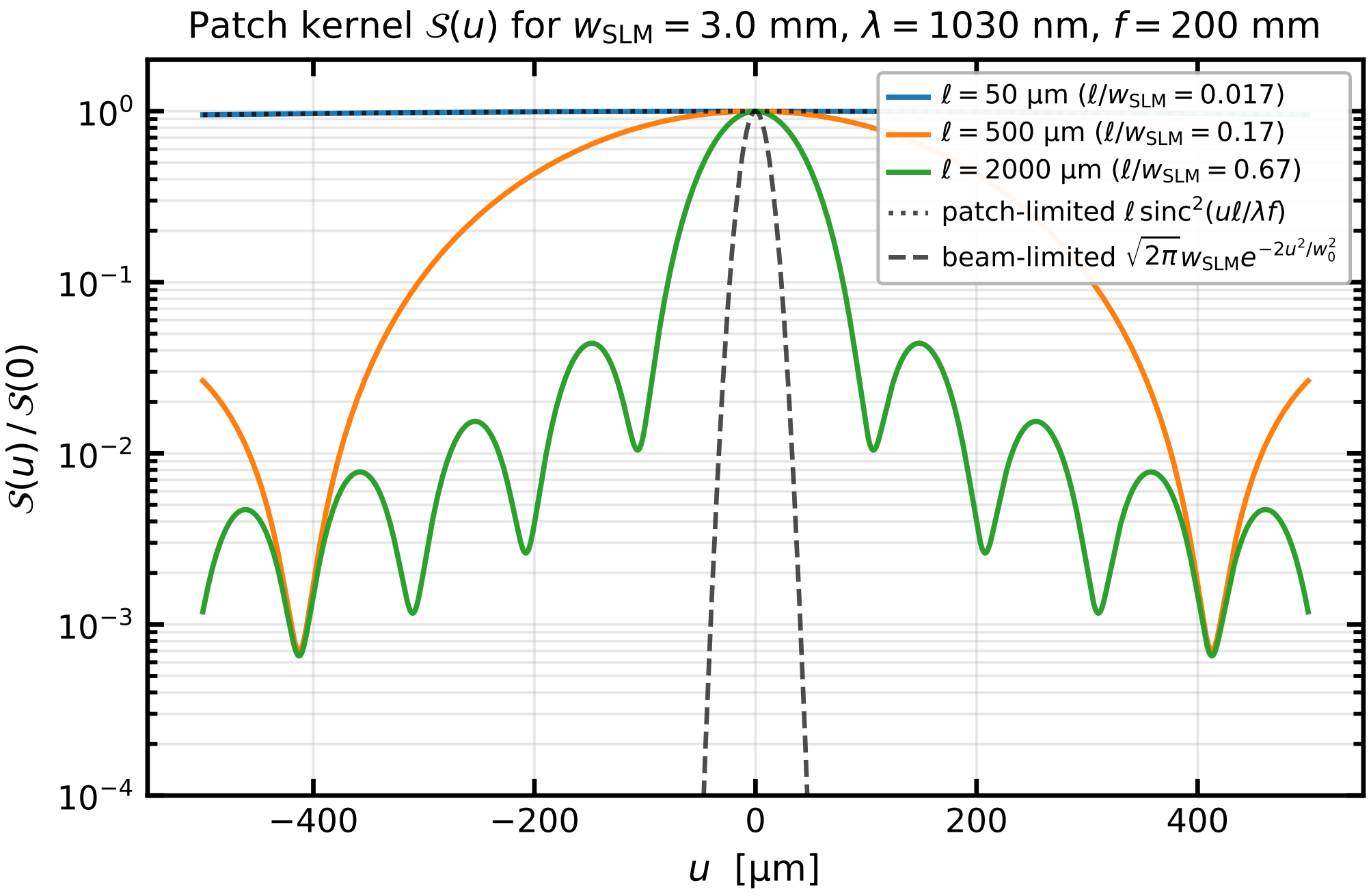

**Figure S3:** Patch kernel $\mathcal{S}(u)$ of Eq. (S24), normalized to $\mathcal{S}(0)$ and plotted against focal-plane position, for three patch sizes spanning the patch-limited to beam-limited crossover ($w_{\mathrm{SLM}} = 3$ mm, $\lambda = 1030$ nm, $f = 200$ mm). At small $\ell/w_{\mathrm{SLM}}$ the kernel is broad and follows the patch-limited sinc$^2$ (dotted). At large $\ell/w_{\mathrm{SLM}}$ it collapses onto the beam-limited Gaussian (dashed) of Eq. (S26).

### 4.3 Cross-channel halos

For the cross-channel piece the midpoint and separation factors decouple in each pair. The midpoint integrals are Gaussian-cosine integrals,

$$\iint |A_{\mathrm{SLM}}|^2 \cos(aX_m + bY_m)\, dX_m\, dY_m = P_{\mathrm{tot}}\, e^{-(a^2+b^2)w_{\mathrm{SLM}}^2/8},$$

so each pair is suppressed by

$$\varepsilon_{AB} = e^{-k_t^2 w_{\mathrm{SLM}}^2/2}, \qquad \varepsilon_{AD} = \varepsilon_{BD} = e^{-(1+n_D^2)\,k_t^2 w_{\mathrm{SLM}}^2/8}. \tag{S27}$$

The separation factor of $C_{AB}$ is unity, so its transform is unshifted and the halo sits at the origin of the deflected frame, midway between the two main foci,

$$\mathcal{M}_{AB}(x,y) = -2\,\xi_A\xi_B\,\varepsilon_{AB}\,\mathcal{S}(x)\,\mathcal{S}(y). \tag{S28}$$

The mixed pairs carry a separation phase, which shifts their transform to the midpoint between the two foci involved,

$$\mathcal{M}_{AD} = -2\,\xi_A\xi_D\,\varepsilon_{AD}\,\mathcal{S}\!\left(x - \tfrac{d}{4}\right)\mathcal{S}\!\left(y - \tfrac{n_D d}{4}\right), \tag{S29}$$

$$\mathcal{M}_{BD} = -2\,\xi_B\xi_D\,\varepsilon_{BD}\,\mathcal{S}\!\left(x + \tfrac{d}{4}\right)\mathcal{S}\!\left(y - \tfrac{n_D d}{4}\right), \tag{S30}$$

and the total cross-channel contribution is

$$\left\langle |\delta E|^2 \right\rangle_{\text{cross}} = \frac{P_{\text{tot}}}{(\lambda f)^2} \left[ \mathcal{M}_{AB} + \mathcal{M}_{AD} + \mathcal{M}_{BD} \right]. \tag{S31}$$

For the experimental parameters, $k_t w_{\text{SLM}} = 4.6$, giving $\varepsilon_{AB} \simeq 3 \times 10^{-5}$ and $\varepsilon_{AD} = \varepsilon_{BD} < 10^{-100}$. The cross-channel contribution is therefore dropped in the main text.

### 4.4 Total focal-plane intensity

Collecting the coherent foci of Eq. (S8), the halos of Eq. (S25) and the cross-channel terms of Eq. (S31),

$$\begin{aligned} I_{\text{tot}}(x,y) \equiv \left\langle |E(x,y)|^2 \right\rangle = I_0 \sum_c \xi_c^2 \, \mathcal{G}(x - x_c,\, y - y_c) \\ + \frac{P_{\text{tot}}}{(\lambda f)^2} \sum_c \xi_c (1 - \xi_c) \, \mathcal{S}(x - x_c) \, \mathcal{S}(y - y_c) \\ + \frac{P_{\text{tot}}}{(\lambda f)^2} \left[ \mathcal{M}_{AB} + \mathcal{M}_{AD} + \mathcal{M}_{BD} \right], \end{aligned} \tag{S32}$$

with $\mathcal{G}(x,y) = e^{-2(x^2+y^2)/w_0^2}$, $\mathcal{S}$ the kernel of Eq. (S24) and $I_0 = 2P_{\text{tot}}/(\pi w_0^2)$. Dropping the cross-channel line recovers the expression quoted in the main text.

## 5 Focal-plane intensity of the unshaped beam

Close to the foci, the measured cuts of Fig. 3 of the main text level off on a pedestal around $-20$ dB, which we attribute to the beam itself, through imperfect beam quality and scattering in the optics upstream. Figure S4 shows the corresponding cuts for the unshaped beam, with a flat mask on the SLM. The same pedestal appears around $-20$ dB close to the focus, together with similar structure in its wings, so it is a property of the beam and of the optics upstream rather than of the stochastic mask.

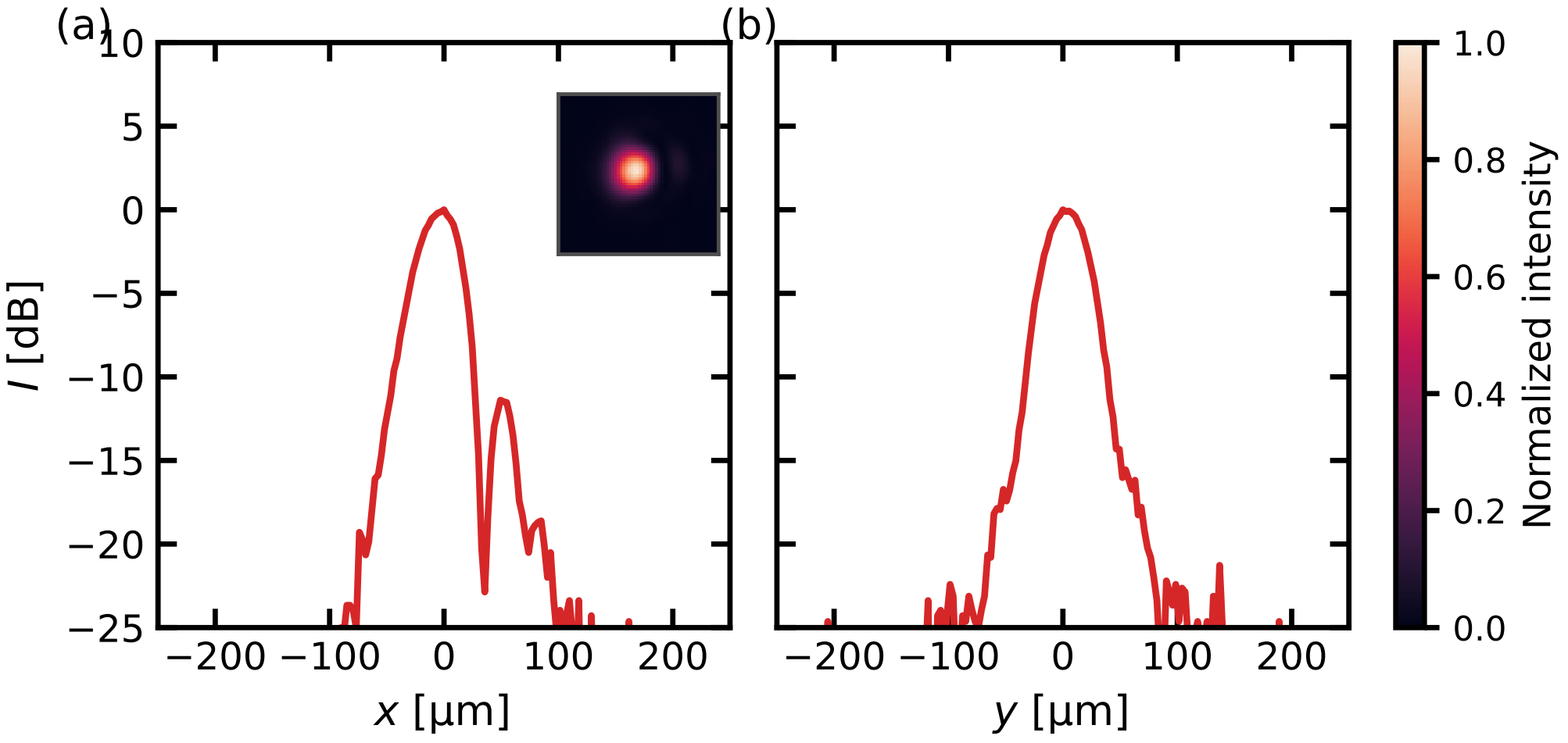

**Figure S4:** Focal-plane intensity cuts of the unshaped beam along (a) $y = 0$ and (b) $x = 0$, on a log scale relative to the peak. Inset: measured focal-plane image.